# The mixture of glycerin with tartrazine: a solution to reversibly increase tissue transparency for *in vitro* quantitative phase imaging

MIKOŁAJ KRYSA[1,*], ANNA CHWASTOWICZ[2,3], MAŁGORZATA LENARCIK[4,5], PAWEŁ MATRYBA[2], PIOTR ZDAŃKOWSKI[1] AND MACIEJ TRUSIAK[1,**]

[1]*Institute of Micromechanics and Photonics, Warsaw University of Technology, Warsaw, Poland*
[2] *Department of Immunology, Medical University of Warsaw, Warsaw, Poland*
[3] *Laboratory of Neurobiology, Nencki Institute of Experimental Biology of the Polish Academy of Sciences, Warsaw, Poland*
[4]*Department of Pathology, Maria Sklodowska-Curie National Research Institute of Oncology, Warsaw, Poland*
[5]*Department of Gastroenterology, Hepatology and Clinical Oncology, Centre of Postgraduate Medical Education, Warsaw, Poland*
*\*mikolaj.krysa@pw.edu.pl*
*\*\*maciej.trusiak@pw.edu.pl*

**Abstract:** Thick tissue sections strongly scatter and absorb light, which limits transmission-based label-free examination via quantitative phase imaging (QPI) modalities. Here we introduce a simple, room-temperature optical clearing medium - glycerol and tartrazine solution (GTS; 60% glycerol, 10% tartrazine) - that increases the transparency of 50–80 μm murine liver and kidney slices while preserving tissue morphology and enabling rapid, label-free quantitative phase imaging. Using Fourier ptychographic microscopy (FPM) and lensless digital holographic microscopy with pixel super-resolution (LDHM-PSR), we demonstrate markedly improved high-throughput visualization of microstructural features after GTS immersion compared with phosphate-buffered saline (PBS). The improvement is confirmed quantitatively by a significant increase in the one-pixel-lag autocorrelation of phase-gradient values, indicating reduced scattering-driven phase artifacts and enhanced structural continuity in reconstructed phase maps. We further show that GTS is stable for months at room temperature, provides instant yet durable clearing that improves over time, and can be removed by brief PBS washing, enabling downstream multimodal analyses on the same specimen. Finally, we benchmark GTS against the state-of-the-art clearing agent Ce3D, highlighting GTS as a low-cost, safe, operationally straightforward alternative with strong potential for routine, high-throughput QPI-based histopathology workflows.

## 1. Introduction

Tissue sections are among the most frequently used sample types for patient diagnosis and biomedical research. Beyond detecting specific structures, such as cell types, organelles, or tumors, they enable the visualization of spatial morphological relationships - information that is crucial for accurate diagnosis and understanding biological mechanisms. Visualization using classical bright-field microscopes is limited to observing structures with high internal intensity contrast If the sample does not exhibit inherent absorption-based variability, it must be stained. Classical chemical staining, however, while it usually maintains sample morphology, often leads to significant changes in the biochemistry of the stained specimen [1,2]. This, in turn, leads to the limitation of the modalities that can be used to image or analyze the sample. Immunostaining, on the other hand, typically does not significantly alter the sample's biochemistry. However, the efficiency of staining can be problematic, and the high-intensity light required to excite the fluorophores may lead to photobleaching, resulting in cell damage and the inability to visualize the sample [3,4]. Moreover, the fluorescent labels have limited visualization capabilities, since they are usually highly specific to target proteins. Furthermore, optimization of an individual staining panel can be labor-intensive and costly. Label-free techniques can address all of those problems [5,6], and Quantitative Phase Imaging (QPI) is among the most promising for examining *in vitro* biological specimens [7–9].

QPI techniques visualize samples by measuring the phase delay introduced as light passes through them, which is directly proportional to the sample's thickness and its refractive index (RI) [10–12]. Because QPI detects phase shifts, it enables label-free investigation of the optical thickness variations of the samples. Two of the most promising techniques in the context of high space-bandwidth product [13] tissue section imaging are Fourier Ptychographic Microscopy (FPM) [14–17] and Lensless Digital Holographic Microscopy [18–22] with Pixel Super-Resolution [23–27] (LDHM-PSR). Both of these techniques offer a large field of view (FOV) with a high (FPM, e.g., 490 nm) to modest (LDHM-PSR, e.g, 1.2 μm) spatial resolution, with the space-bandwidth product generally being much higher than that of other QPI modalities [17,28,29]. However, since both of these techniques are transmission-based, imaging thick (> 10 microns), highly RI-variant, highly light-absorbing, and/or strongly scattering specimens remains a challenge [30,31]. Unfortunately, tissue samples (especially thick ones) substantially scatter and absorb the light passing through them [32], limiting the use of those two techniques in this context. Some researchers have attempted to overcome these limitations by using optical clearing agents (OCAs) to enable imaging of the sample using QPIs. One attempt involved using a glycerol solution in saline (70% v/v) to minimize light scattering and enable deep imaging of porcine skin tissue using digital holographic microscopy [33]. Another used the simplified CLARITY protocol to clear and stain neurons in 200 μm brain tissue samples for LDHM-PSR imaging [34]. Apart from that, the whole zebrafish larvae imaging with Optical Diffraction Tomography (ODT) was performed after clearing with a benzyl alcohol and benzyl benzoate (BABB) mixture [35]. Moreover, live imaging using ODT of *Candida albicans* and *Caenorhabditis elegans* was possible after RI matching with iodixanol [36]. Although effective in the settings shown in the articles, all those clearing solutions have certain drawbacks – they might be of low efficiency (glycerol) [37], induce substantial chemical changes in the sample, and therefore prevent the ability to use other analysis modalities (CLARITY and BABB), have a laborious and long protocol (CLARITY), have a high cost which limits their suitability for a clinical use (such as iodixanol) or pose a health risk for a personnel using it (CLARITY and BABB) due to toxic substances used throughout the protocols.

Besides the aforementioned OCAs that have already been applied in QPI, there are other promising candidates, such as Ce3D, which has been used as a clearing agent in fluorescence imaging and is known not to induce significant chemical changes in the sample, but it has not yet been explored in QPI applications. This solution increases the sample's transparency mainly by matching the RI of the medium to that of the tissue, which reduces scattering and absorbance [38]. Apart from fluorescence imaging, it has been demonstrated to be effective for RNA detection and shows promise in the context of QPI. Unfortunately, it has two main drawbacks for use in clinical settings - one of the components (N-methylacetamide) has a potentially toxic effect on human reproduction, and another (HistoDenz) is expensive [38]. There is also a drawback - Ce3D contains Triton X-100, which is incompatible with Liquid Chromatography Mass Spectrometry due to ion suppression [39]. This limits the samples' subsequent use for that analysis.

Recently, the chromophores with strong absorption in a specific spectral band were shown to increase the RI of an aqueous medium in other-than-absorbed wavelengths [40]. The most promising of the chromophores - tartrazine, a common food dye, was shown to induce transparency of the living murine skin, enabling visualization of internal organs with white light, muscles with second harmonic generation, and cholinergic neurons in the gut of the genetically modified mouse [40]. After this seminal work, tartrazine was shown to be effective in label-free visualization of skin depth structure in living mouse using Optical Coherence Tomography (OCT) [41], improving the visualization of *ex vivo* porcine eye depth structure using OCT [42], enabling visualization of NIR probes in tissue phantoms using fluorescence lifetime imaging [43], and improving the visualization of blood vessels in the ear, abdomen, head, and leg using photoacoustic microscopy [44–46]. These successes in rendering the tissue transparency and enabling the visualization of the internal structure using label-free techniques lead to the hypothesis that it might also be successful in enabling the visualization of thick tissue slices using *in vitro* QPI. However, the most effective concentration to optically clear the tissue with tartrazine is ~30%. Unfortunately, such a high concentration precipitates at room temperature, rendering the sample opaque [44]. Decreasing the tartrazine concentration alone is not a solution, as its effectiveness is directly proportional to concentration [40].

To overcome this limitation, we introduce a Glycerol and Tartrazine Solution (GTS) for the optical clearing of the thick tissue slices for *in vitro* QPI at room temperature. This solution does not cause chemical or morphological alterations, is reversible, and thus enables usage for subsequent analyses (with e.g. H&E staining). Moreover, it is very straightforward to use, operationally inexpensive, clears the sample instantly, is non-toxic, and has a long shelf life, which enables its routine use in laboratory and clinical environments. Most importantly, it effectively renders the sample transparent, enabling the visualization of microstructure in thick tissue slices using high-throughput label-free histopathological scanning with FPM and LDHM-PSR frameworks.

In the following sections, we demonstrate the effectiveness of the GTS on 50- and 80 μm mouse liver and 80 μm mouse kidney slices. We present both the visual (qualitative) and numerical (quantitative) effects of GTS clearing. Moreover, we examine the non-invasiveness with respect to tissue morphology, the stability of GTS, and the reversibility of its clearing. We also compare GTS with other promising OCA (Ce3D) in terms of its ability to display sample internal microstructure and solution dynamics. Finally, we discuss the limitations of the GTS, providing a comprehensive review of the GTS's potential in routine laboratory or clinical settings.

## 2. Results and Discussion

### 2.1 Optimizing the concentration of glycerol and tartrazine

Since tartrazine precipitates in high concentrations, the concentration of glycerin and tartrazine had to be optimized. The tested tartrazine concentrations were 10%, 20%, and 30%, while the tested glycerol concentrations were 20%, 40%, 60%, and 75%. The test was performed in each combination (except 75% glycerin with 30% tartrazine, as this combination was impossible). All the solutions with 20% and 30% tartrazine concentrations precipitated at room temperature. Additionally, 10% tartrazine did not dissolve in 75% glycerol. The highest glycerol concentration with 10% tartrazine was therefore 60% glycerol, and this solution was chosen as an optimized concentration. The highest possible glycerol concentration was selected because glycerol has a higher RI than water. The higher the concentration, the higher the RI, and therefore the higher the transparency of the high-RI samples, such as tissues, due to reduced light scattering and refraction [47,48].

### 2.2 Glycerol and tartrazine solution enable phase imaging of thick tissue slices

The optical clearing properties of GTS were showcased using two large FOV QPI methods: FPM (Fig. 1A) and LDHM-PSR (Fig. 2A). They were chosen because one of them uses a low-coherence light (FPM utilizes low-coherent LEDs), and the second uses light of high coherence (LDHM-PSR utilizes a laser). GTS proved to be highly effective in both QPI modalities compared to Phosphate Buffered Saline (PBS), as shown in Figs. 1A and Fig. 2A.

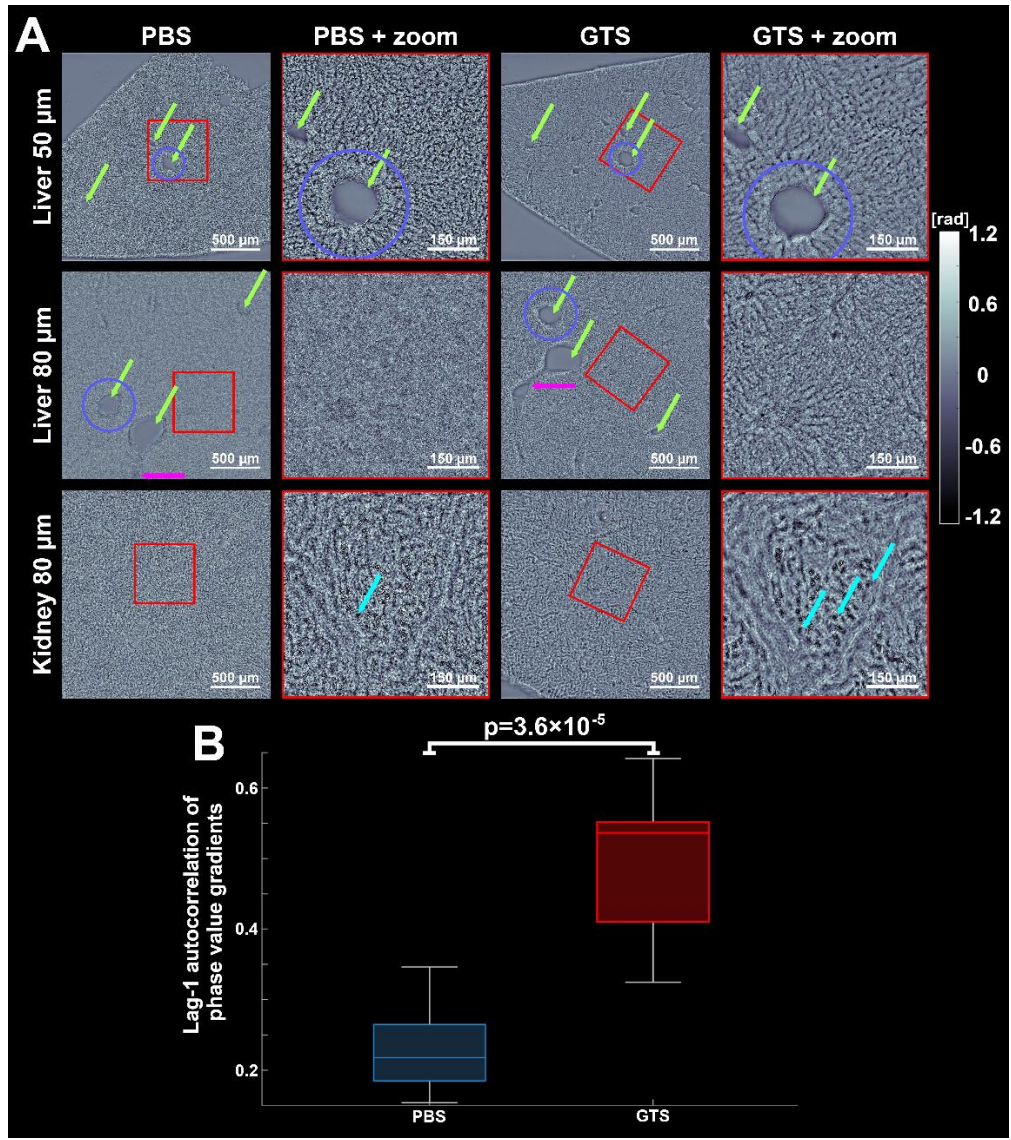

Fig. 1. The image (A) and quantitative (B) results obtained from Fourier Ptychographic Microscopy of the liver (50 μm and 80 μm thick) and kidney (80 μm thick). The images (A) show the reconstructed phase from the samples in PBS and in Glycerol and Tartrazine Solution (GTS). The red rectangles mark the zoomed region. The corresponding areas are shown in the zoomed images. The color bar shows the phase values. Green arrows indicate central veins, magenta arrows indicate hepatic sinusoids, cyan arrows indicate glomeruli, and blue circles indicate hepatic lobules. Only a few structures are shown in the images. The quantitative results (B) are box plots of autocorrelation of phase value gradients at a lag of one pixel of the sample before (PBS) and after optical clearing (GTS). The value above graph indicates the p-value obtained with a paired t-test. The analysis in PBS and GTS was carried out on the same slice.

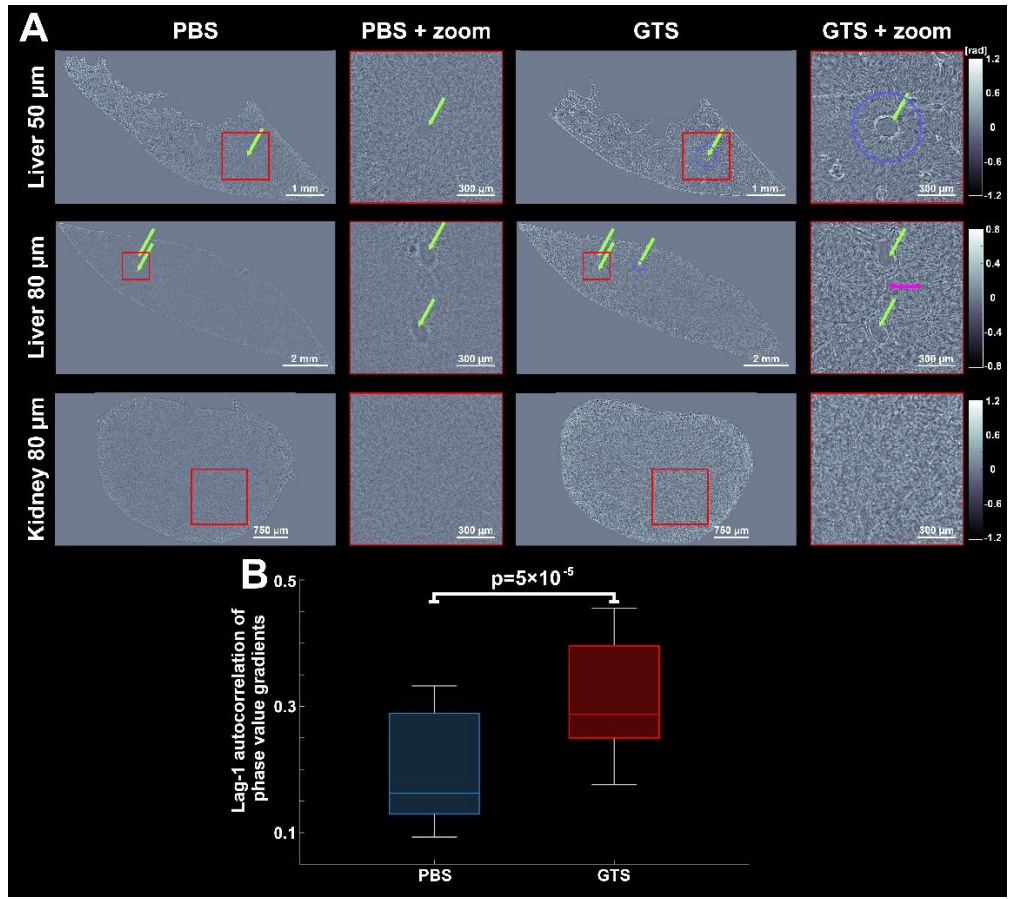

Fig. 2. The image (A) and quantitative (B) results obtained from Lensless Digital Holographic Microscopy with Pixel Super Resolution of the liver (50 µm and 80 µm thick) and kidney (80 µm thick). The images (A) show the reconstructed phase from the samples in PBS and in Glycerol and Tartrazine Solution (GTS). The red rectangles mark the zoomed region. The same areas are shown in the zoomed images. The color bar shows the phase values. Green arrows indicate central veins, magenta arrows indicate hepatic sinusoids, and blue circles indicate hepatic lobules. Only a few structures are shown in the images. The quantitative results (B) are a box plot of the autocorrelation of phase value gradients at a lag of one pixel for the sample before (PBS) and after optical clearing (GTS). The value above the graph indicates the p-value obtained with a paired t-test. The analysis in PBS and GTS was carried out on the same slice.

In the case of the FPM (Fig. 1A), while PBS-covered 50 µm liver slices appear satisfactory at first glance - it enables visualization of tissue architecture and identification of hepatic lobule structure - a closer examination reveals phase artifacts characteristic of highly scattering objects with low transparency. Using GTS as a medium enabled the visualization of the 50 µm liver samples with minimized scattering phase artifacts. Similar effects are observed in the case of an 80 µm kidney, however, the slight scattering phase artifacts persist even after using GTS. Unlike previous examples, PBS alone does not enable the visualization of any microscopic structures, except for blood vessels, in 80 µm liver slices. In contrast, GTS enables this visualization even without scattering phase artifacts (see also Fig. S3 in the Supplementary Document 1 for zoomed-in comparison of amplitude and phase reconstructions).

In the case of LDHM-PSR (Fig. 2A), details are largely invisible in the PBS-covered 50 µm liver samples, and no details are visible in PBS-covered 80 µm samples (liver and kidney). The use of the GTS enables the visualization of microstructural details in all cases, although overlapping cell layers hinder the distinction of finer details, especially in 80 µm slices. Two other slices of each type were tested on FPM and LDHM-PSR to test whether the optical clearing properties of the GTS are consistent. The results (Figs. S1 and S2 in Supplementary Document 1) confirm this.

Both in FPM and LDHM-PSR, hepatic lobules were clearly delineated, displaying a polygonal shape with a centrally located central vein, radially arranged hepatocyte plates, and interspersed hepatic sinusoids in GTS-treated liver samples. Peripheral structures consistent with portal triads were also observable. In the GTS-treated kidney tissue, numerous cross-sections of renal tubules and blood vessels, which form glomeruli, were evident. Using GTS for the optical clearing of thick tissue slices might therefore enable the interpretation of histological imaging results in biomedical diagnostics.

Apart from a visual confirmation of the GTS effectiveness, we have taken advantage of the quantitative nature of the QPI and examined the autocorrelation of phase value gradients at a lag of one pixel. This metric is calculated by measuring the differences between horizontally and vertically adjacent pixel pairs, and then computing the Pearson correlation between consecutive gradient values in the sequence. In the context of QPI, this metric refers to the degree of linearity and smoothness in the trend of spatial phase value change. The higher the number, the more linear and smooth the changes are. If the value is closer to zero, the more random noise changes are present [49]. The analysis of this metric from 9 samples imaged with FPM and LDHM-PSR shows that the GTS significantly increases the correlation between subsequent pixels in both techniques. The mean correlation increased from 0.23 in PBS to 0.49 in GTS in the FPM imaging (Fig. 1B) and from 0.2 to 0.31 in LDHM-PSR (Fig. 2B). This indicates a substantial increase in the smoothness of the phase changes, thereby improved internal structure visualization. The larger increase observed in the FPM cases can be attributed to the low-coherence LED light source used, which physically limits coherent artifacts and noise. On the other hand, in the LDHM-PSR, the laser was used, producing more coherent noise.

### 2.3 Glycerol and tartrazine solution minimally changes the morphology of the tissues, it is stable, and its clearing is reversible

To evaluate the solution's impact on the overall sample morphology, the change in area was calculated for samples covered with PBS, followed by those covered with GTS. The mean shrinkage of kidney slices was 9.37% (SD=6.14%), while the mean area of liver slices almost did not change (it increased by 0.7% with SD=4.67%). One of the factors that drives the shrinkage or swelling of tissues is the difference between the solution in which the tissue is submerged and the tissue's own osmolality [50]. Mammalian tissue osmolality is typically ~300 mOsm, while calculated GTS osmolality is 6800 mOsm. Considering this, the slight change in the area is quite surprising – water should be driven out of the sample based on this factor. It was, however, not observed – the tissue shrinkage is low in the kidney and none in the liver. A possible explanation is that the samples are relatively thin and have a large area-to-volume ratio, which enables the rapid replacement of the driven-out water with glycerol and tartrazine, and thus faster reaching of osmotic equilibrium. [51]. This shortens the shrinkage period and therefore limits it [52]. The difference in shrinkage between the kidney and the liver may be due to the variation in the amount of extracellular matrix (ECM) between the two organs – it is higher in the kidney and lower in the liver [53,54]. A higher amount of ECM might lead to a limitation in the speed of glycerol influx, while limiting water efflux to a lesser degree, resulting in higher tissue shrinkage in samples with higher ECM content [55]. Moreover, it is worth considering that the number of slices analyzed to calculate the area change was small – 3 in the kidneys (80 µm thick) and 6 in the liver (50 µm and 80 µm thick). Therefore, the exact percentage of area change may differ if more samples were used, although the tendency should be similar to the one presented in this paper.

Another property of the GTS is its stability. Once prepared, GTS was successfully used to clear tissues for 5 months without compromising its optical clearing properties. Moreover, it was kept at room temperature throughout this period, and its clearing properties remained unchanged, which further supports its use in everyday practice.

Apart from the stability of the solution, the sample submerged in the GTS is also stable. To test this kind of stability, the sample was first imaged in PBS. Afterwards, the same sample was submerged in GTS and left in this solution for three days. Every day, the image was captured using FPM (Fig. 3A). Furthermore, the autocorrelations of phase value gradients at a lag of one pixel were extracted from the images (Fig. 3C). This statistic, along with the images, revealed a gradual increase in microstructure visibility over time (Fig. 3A, C).

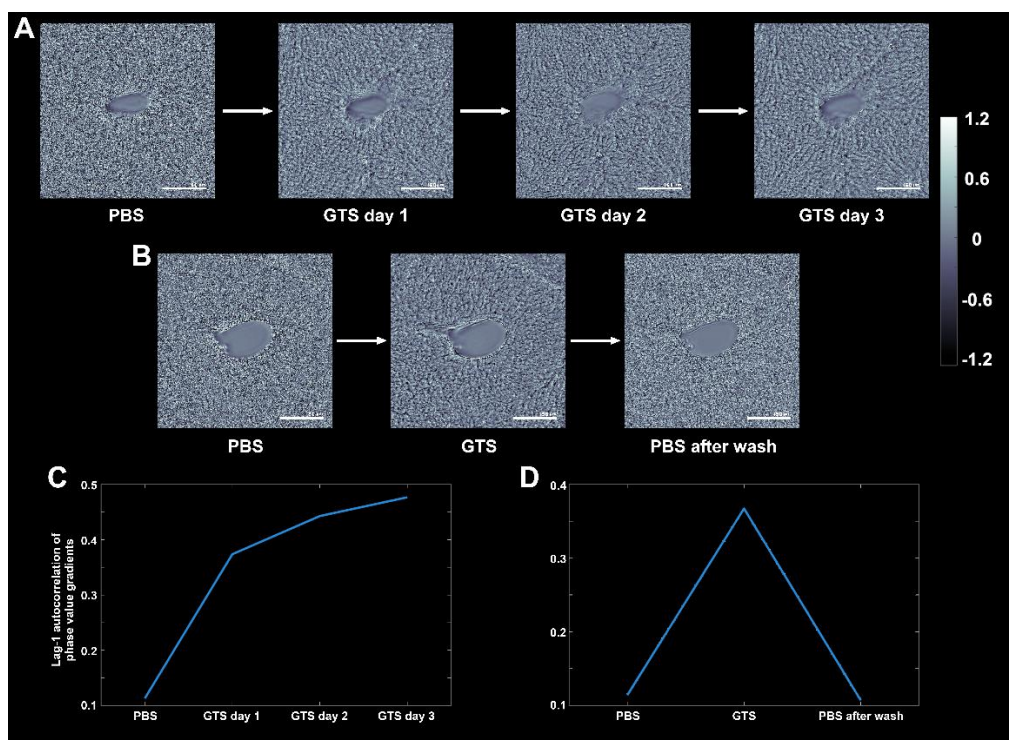

Fig. 3. The presentation of the stability of the sample visualization submerged in the glycerol and tartrazine solution (GTS) over days (A, C), and presentation of the reversible GTS clearing effect (B, D). The stability was investigated using 80 μm liver tissue over a period of 3 days, and is shown in the images (A) and in the autocorrelation values of phase gradients at a lag of one pixel (C). The reversible effect of GTS clearing was also shown on 80 μm liver slices. The sample was first submerged in PBS, then in GTS, and subsequently washed for 15 minutes with PBS. It was imaged after each of these steps, and the results are shown as reconstructed phase images (B) and numerically as the autocorrelation of phase value gradients at a lag of one pixel (D). Each analysis was carried out on a single slice, and the slices were imaged using Fourier Ptychographic Microscopy. The color bar shows the phase values of the images. Scale bar indicates 150 μm.

GTS is composed of glycerol and tartrazine, both of which are hydrophilic and do not bind directly to the tissue. This means that they should be easily cleansed from the tissue after usage. Indeed, this effect was observed in our study (Figs. 3B, 3D and S4). We observed that washing for 15 minutes in PBS reverses the clearing effect of GTS, indicating that most of the GTS components are removed. It was presented both in the phase (Fig. 3B) and amplitude (Fig. S4 in the Supplementary Document 1) images, as well as quantitatively using the autocorrelation of phase value gradients at a lag of one pixel, which yielded the same value after PBS washing as before GTS application (Fig. 3D). Similarly, reverse clearing effects were observed in studies using single components (glycerin and tartrazine separately) [40,56]. Removing GTS components enables the same tissue to be used for other analyses, e.g., another imaging modality or biochemical analysis. This, in turn, might provide a deeper insight into the correlation between sample morphology and biochemistry.

### 2.4 The comparison of glycerol and tartrazine solution dynamics and clearing properties with Ce3D optical clearing agent

To test the dynamics of GTS, a video recording of the clearing process of an 80 µm liver slice was employed. Moreover, to compare its dynamics with another mild optical clearing agent, the same analysis was also performed for Ce3D, which acts as a fast RI-matching solution. The dynamic change is shown in Visualization 1 and in Figs. 4A and 4B.

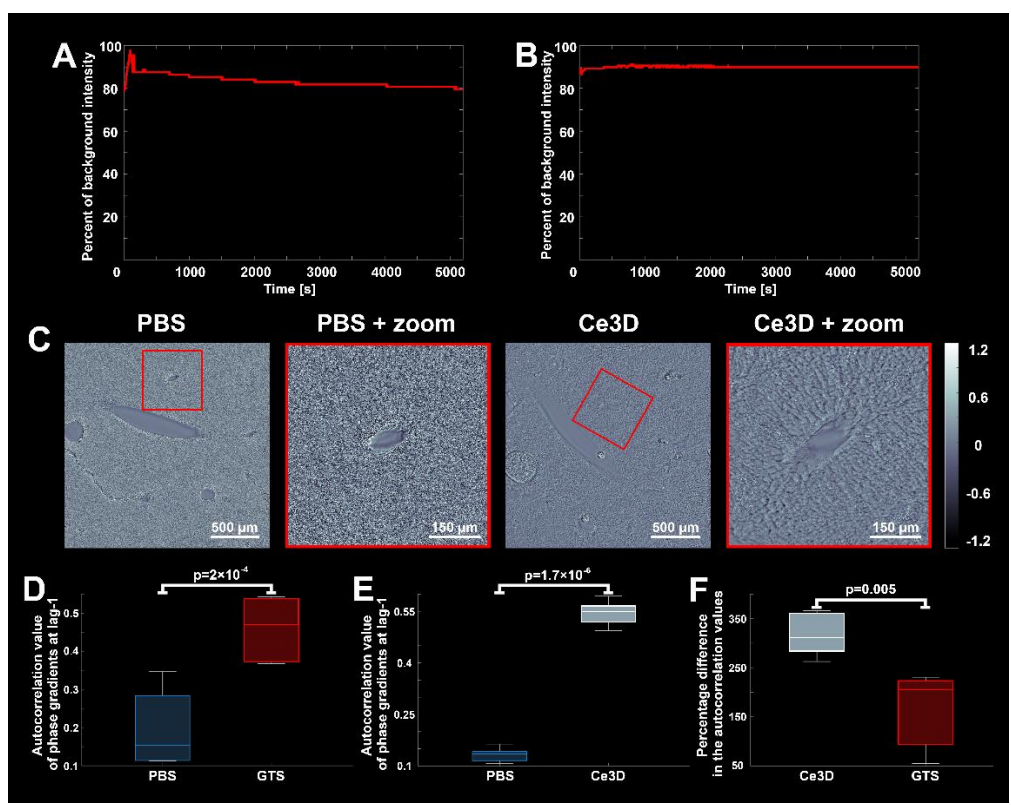

Fig. 4. The comparison of the dynamics and clearing properties of the glycerol and tartrazine solution (GTS) with Ce3D optical clearing agent. The dynamics of clearing were calculated as the percentage change in median pixel intensity compared to the background over time. The investigation was conducted after the application of glycerol and tartrazine solution (A) and Ce3D solution (B). The pixels taken for the analysis were acquired only from the sample area. The image was acquired at a rate of 1 Hz for 5200 seconds, starting from the moment the solution reached the sample. The clearing properties of Ce3D compared to PBS in the 80 µm liver slices are shown in the image (C) and as the autocorrelation of phase value gradients at a lag of one pixel (E). The same slice thickness was used to compare PBS with GTS (D). The percentage difference between PBS and the optical clearing agent was used to represent the clearing effectiveness of Ce3D and GTS (F). The values above the box plots (D, E, F) represent the p-values of the calculated statistical significance.

The dynamics of solutions tested were quite similar – they reached nearly maximum clearance immediately after contact with the solution, however, there were slight differences. GTS had a transparency of 78% of the background in the first second of the analysis. The transparency then increased dramatically, reaching nearly 98% of the background intensity, before decreasing to ~88% after 130 seconds. Afterwards, the intensity then decreased slowly, reaching ~80% after 5076 seconds and stayed at this level for the remainder of the analysis (Fig. 4A). In contrast, Ce3D did not exhibit a rapid intensity increase during the first 130 seconds. It caused a ~89% background intensity of the sample in the first second of the measurement, then it had a slight drop to 87%, but it was back to 89% after approximately 85 seconds and remained almost constant for the rest of the analysis (Fig. 4 B). Despite slight differences, this analysis shows quite similar dynamics for both solutions.

Apart from the dynamics analysis, the clearing properties in the context of quantitative phase imaging of GTS were compared to those of Ce3D. Similar to GTS, Ce3D also enables visualization of the 80 µm liver microstructure (Fig. 4C). However, to compare the clearing process numerically, 10 samples of 80 µm liver samples were FMP imaged in PBS, and then 5 were imaged in GTS and 5 in Ce3D. Then, the autocorrelations of phase value gradients at a lag of one pixel were extracted from the PBS, GTS, and Ce3D submerged samples. The paired t-tests were performed on PBS-OCA pairs of the samples, and they showed significantly increased autocorrelation values after the use of both OCAs (Fig. 4D and E). Moreover, the percentage difference in the autocorrelation values of the gradients at a lag of 1 for OCA-treated images compared to PBS was calculated for each sample. Since the distribution of the change was normal for both the GTS and Ce3D groups, a t-test was performed. The results showed that the percentage difference from the PBS was significantly higher in Ce3D (mean = 318%) compared to GTS (mean = 164%), indicating a more pronounced clearing effect and a greater ability to visualize the liver microstructure in the phase images after Ce3D treatment compared to GTS. The detailed values of the differences between the groups are shown in Fig. 4F. However, it is worth noting that Ce3D has a lower clinical setting and multimodal analysis potential compared to GTS, due to its higher cost, toxicity, and the presence of Triton X-100 [38].

*2.5 Limitations of glycerol and tartrazine solution in clinical settings*

Although GTS has many advantages as an optical clearing agent in clinical settings (as outlined above), it does face some limitations. One is that it is very easy to stain laboratory equipment while GTS is used, due to the intrinsic property of the tartrazine – it is a dye. Of course, due to its hydrophilicity, it is relatively easy to wash off, however, this might be quite inconvenient.

Moreover, tartrazine exhibits high absorption in the 200-290 nm and 350-490 nm regions, which limits its use in potential fluorescence microscopy settings that utilize excitation or emission wavelengths in these spectral regions [44].

## 3. Conclusions

This study demonstrated that GTS facilitates the QPI of thick (50 µm and 80 µm) animal tissue samples, as showcased using FPM and LDHM-PSR. GTS proved to be a good candidate for a standard clinical use due to its powerful optical clearing properties, replicability, minimal tissue volume change, minimal toxicity, long "shelf life", instantaneous but long-lasting clearing, and a very low cost compared to other clearing methods suitable for QPI. Moreover, due to the ability to wash the GTS components out, the same sample that was previously cleared with GTS and imaged with QPI can be used for other analyses.

## 4. Materials and Methods

*4.1 Preparation of murine sample slices*

Murine kidneys and livers were obtained from inbred C57BL/6J mice (bred at the Center for Experimental Medicine in Białystok and the M. Mossakowski Medical Research Center in Warsaw). The animals were provided a controlled environment (temperature 24°C, 12/12 light/dark cycle) with ad libitum access to water and feed. The animals were euthanized by cervical dislocation following exposure to isoflurane (FDG9623, Baxter). Isolated tissues were fixed for 24 h in a 4% paraformaldehyde solution (BD Cytofix/Cytoperm, No. 554722, BD Biosciences) at 4°C before being used in experiments. All procedures were conducted in accordance with the Directive of the European Parliament and Council No. 2010/63/EU on the protection of animals used for scientific purposes under a protocol approved by the II Local Ethical Committee for Experiments on Animals in Warsaw, Poland (WAW2/091/2024).

The fixed organs were embedded in a 3% (w/v) agarose solution (Sigma–Aldrich) in water and subsequently sectioned to the desired thickness using a Vibratome (VT1000S, Leica). The sections were stored in 1× PBS (Sigma–Aldrich) supplemented with 0.05% sodium azide (NaN3, Sigma–Aldrich).

### 4.2 Preparation of glycerol and tartrazine solution

The final optical clearing agent used in this study comprised 10% tartrazine (≥85% purity, Pol-Aura Sp. z o.o.), 60% glycerol (≥98.5% purity, Chempur), and ~28,24% demineralized water. The rest of the solution was tartrazine impurities. The concentration of the GTS was optimized to render the tissues transparent without precipitation at room temperature. The optimization process is described in the Results section.

### 4.3 Preparation of Ce3D solution

Ce3D was prepared according to a published protocol [57]. In brief, a 40% (v/v) solution of N-methylacetamide (M26305-100G, Sigma–Aldrich) in PBS was prepared and used to dissolve HistoDenz (D2158-100G, Sigma–Aldrich) to 86% (w/v) at 37°C. After complete dissolution, achieved in 5–6 h, Triton X-100 (0.1%, v/v, Sigma–Aldrich) and 1-thioglycerol (0.5%, v/v, M1753-100ML, Sigma–Aldrich) were added to the solution.

### 4.4 Quantitative Phase Imaging systems

Samples were imaged in two QPI systems: FPM and LDHM-PSR. The FPM system was a typical brightfield microscope (Nikon Eclipse Ei) with a 4x 0.2 NA objective (Nikon Plan Apo λD) and a 2.74 μm camera pixel size (Basler a2A5320-23umBAS CMOS camera). However, the condenser was removed from a microscope, and a light source was replaced with a 7x7 LED array. The array was placed ~70 mm below the sample plane. Red LEDs (centered at 625 nm, 8.1 mm apart from each other) were used in this study. Red light was used because of the high absorption of blue light and moderate absorption of green light by tartrazine. All 49 images, illuminated from different angles, were acquired and used to reconstruct the resolution-increased phase image using the Quasi-Newton algorithm [58] (synthetic NA≈0.54). The images were acquired and reconstructed using MATLAB.

LDHM-PSR setup consisted of a red laser with a 632.8 nm central wavelength, a beam collimator, and a 2.4 μm pixel-sized camera (Allied Vision Alvium 1800 U-2050c) connected to the motorized stage that moved away ~50 μm in the $z$ axis after every image acquisition (axial scanning was employed for the phase retrieval and achieving sub-pixel resolution). Twenty images were acquired for a single image reconstruction. The method for phase retrieval and achieving sub-pixel resolution was previously described by our group [59]. The LDHM-PSR system was controlled, and images were reconstructed using MATLAB.

### 4.5 Experiment workflow

The slices were placed on a standard glass slide, a drop of PBS was added to the sample, and then a coverslip was applied. The sample was then imaged with both systems (FPM and LDHM-PSR). After imaging, the coverslip and PBS were removed with a dust-free wipe, and one or two drops of GTS, followed by a coverslip, were placed on the sample. The sample was then imaged again using FPM and LDHM-PSR. Afterward, the samples were imaged as mentioned in previous sections.

### 4.6 Phase image properties and statistical analysis

The autocorrelations of phase value gradients at a lag of one pixel were extracted from each phase image. The formula for this is presented in Eq. 1.

$$r_1 = \frac{\sum_{i=1}^{N-1}(g_i - \bar{g})(g_{i+1} - \bar{g})}{\sum_{i=1}^{N}(g_i - \bar{g})^2} \tag{1}$$

Where: $g_i$ is the $i$-th intensity gradient value, $\bar{g}$ is the mean of all $g_i$ values, N is the number of computed gradient values. The autocorrelations of phase value gradients at a lag of one pixel were calculated only from the sample area (excluding the background and holes). The differences of those values before and after the use of GTS were calculated, and the distribution tests (Lilliefors) were performed separately for the FPM and LDHM-PSR investigated samples. If the distribution was normal, a paired t-test was performed. Otherwise, the Wilcoxon signed-rank test was used. Moreover, the box plots showing the autocorrelations of phase value gradients at a lag of one pixel for the FPM and LDHM-PSR were created. All of the above-mentioned analyses were performed using MATLAB.

### 4.7 Calculating the glycerol and tartrazine solution on the sample area

The areas of the samples were calculated from all the samples imaged with LDHM-PSR (three samples of each type). The area was first calculated for the PBS-covered samples and then for the same samples covered with GTS. The area calculation was carried out using ImageJ software (1.54 p).

### 4.8 Investigating the stability of the clearing effect over days

The stability of the sample clearing was tested at three timepoints: immediately after the GTS coverage, after 24 hours, and after 48 hours. The sample used for this purpose was an 80 μm liver section. The sample was first imaged in PBS. Afterwards, the same sample was submerged in GTS and left in this solution for three days. Every day, imaging was performed using FPM.

### 4.9 Investigating the reversible clearing effect

An eighty-micrometer liver slice was placed on the standard glass slide in a drop of PBS, covered with a coverslip, and imaged using FPM. Afterwards, the coverslip and PBS were removed, and one or two drops of GTS, followed by a coverslip, were placed on the sample and imaged again. Then, the coverslip was removed, the slice was transferred to the Petri dish, and washed with PBS. Two minutes afterwards, the PBS was disposed and a fresh PBS was added. It was there for another 13.5 minutes and was gently mixed throughout the process. Then the sample was again placed on a glass slide, a drop of PBS was placed on it, followed by a coverslip. Again, it was imaged with FPM and reconstructed as mentioned in Section 2.3. Afterwards, the autocorrelations of phase value gradients at a lag of one pixel were extracted from the images, and the graph from those values was prepared using MATLAB.

### 4.10   The dynamics of the clearing effect and comparison to Ce3D clearing agent

To investigate the dynamics of the clearing effect, 80 μm liver sections were placed on a standard glass slide and covered with a coverslip. The recording, using a brightfield microscope (the same system as the one used for FPM imaging) with a one-frame-per-second frequency, was turned on, and then the GTS was applied near the coverslip. Due to the capillary action, the solution spread uniformly over the sample. The recording lasted for 5200 seconds from the first contact of the sample with GTS. The analysis was based on the proportion of median light intensity of the sample compared to the background. Only the sample pixels (excluding the background) were utilized, and the selected pixels changed dynamically as the GTS reached the sample, as only the pixels of the sample that were already GTS-covered were used. The percentage of the sample's intensity compared to the background was calculated at each video frame to create the dynamics of the clearing effect. The same procedure was used for a Ce3D solution to compare its dynamics with those of GTS [38].

Apart from dynamic analysis, to compare the effectiveness of the clearing between the GTS and Ce3D, five 80 μm liver samples were used to analyze the effectiveness of the Ce3D clearing process. The procedure was similar to those described previously. Briefly, the samples, covered with PBS and a coverslip, were analyzed using FPM. Then, the coverslip was removed, the PBS was discarded, and Ce3D was placed on the sample. The sample was then covered with the coverslip, and it was analyzed again using FPM. The statistical analysis, as described in Section 2.5, was performed on these results, as well as on five results from 80 μm GTS-treated liver samples for comparison. Apart from statistical analysis, another statistical analysis was performed on the calculated percentage difference of the autocorrelation value of phase gradients at a lag of one pixel between PBS-covered and OCA-covered samples. Since the differences had a normal distribution, a t-test was used to evaluate the significance of the changes between the GTS and Ce3D groups.

**Funding.** National Science Centre, Poland (SONATA 2020/39/D/ST7/03236); The National Centre for Research and Development, Poland (LIDER14/0329/2023; INTENCITY WPC3/2022/47/INTENCITY/2024). The research was conducted on devices cofounded by the Warsaw University of Technology within the Excellence Initiative: Research University (IDUB) programme.

**Acknowledgment.** Authors thank Mikołaj Rogalski and Piotr Arcab for fruitful discussions.

**Disclosure**. The authors declare no conflicts of interest.

**Data availability.** The data underlying the results presented in this paper are available upon reasonable request.

**Supplementary materials.** See Supplementary Document 1 and Visualization 1 for supporting content.

**References**

1. S. T. Rajan and N. Malathi, "Health hazards of xylene: A literature review," J. Clin. Diagnostic Res. **8**, 271–274 (2014).
2. T. Shiomi, A. Eichinger, and L. Chiriboga, "Hematoxylin and Eosin staining of PhenoCycler® Fusion flow cell slides," J. Histotechnol. **46**, 203–206 (2023).
3. J. Icha, M. Weber, J. C. Waters, and C. Norden, "Phototoxicity in live fluorescence microscopy, and how to avoid it," BioEssays **39**, 1–15 (2017).
4. E. Gómez-De-Mariscal, M. Del Rosario, J. W. Pylvänäinen, G. Jacquemet, and R. Henriques, "Harnessing artificial intelligence to reduce phototoxicity in live imaging," J. Cell Sci. **137**, (2024).
5. B. Ghosh and K. Agarwal, "Viewing life without labels under optical microscopes," Commun. Biol. **6**, 559 (2023).
6. N. T. Shaked, S. A. Boppart, L. V. Wang, and J. Popp, "Label-free biomedical optical imaging," Nat. Photonics **17**, 1031–1041 (2023).
7. Y. K. Park, C. Depeursinge, and G. Popescu, "Quantitative phase imaging in biomedicine," Nat. Photonics **12**, 578–589 (2018).
8. T. L. Nguyen, S. Pradeep, R. L. Judson-Torres, J. Reed, M. A. Teitell, and T. A. Zangle, "Quantitative Phase Imaging: Recent Advances and Expanding Potential in Biomedicine," ACS Nano **16**, 11516–11544 (2022).
9. Z. Huang and L. Cao, "Quantitative phase imaging based on holography: trends and new perspectives," Light Sci. Appl. **13**, (2024).
10. J. Zhou, Y. Jin, L. Lu, S. Zhou, H. Ullah, J. Sun, Q. Chen, R. Ye, J. Li, and C. Zuo, "Deep Learning-Enabled Pixel-Super-Resolved Quantitative Phase Microscopy from Single-Shot Aliased Intensity Measurement," Laser Photon. Rev. **18**, 1–16 (2024).
11. Q. Shen, J. Sun, S. Zhou, Y. Fan, Z. Li, Q. Chen, M. Trusiak, M. Kujawinska, and C. Zuo, "Iterative Kramers–Kronig method for non-interferometric quantitative phase imaging: beyond the first-order Born and Rytov approximations," Opt. Lett. **50**, 1144 (2025).
12. X. Ou, R. Horstmeyer, C. Yang, and G. Zheng, "Quantitative phase imaging via Fourier ptychographic microscopy," Opt. Lett. **38**, 4845 (2013).
13. J. Park, D. J. Brady, G. Zheng, L. Tian, and L. Gao, "Review of bio-optical imaging systems with a high space-bandwidth product," Adv. Photonics **3**, 1–18 (2021).
14. G. Zheng, R. Horstmeyer, and C. Yang, "Wide-field, high-resolution Fourier ptychographic microscopy," Nat. Photonics **7**, 739–745 (2013).
15. P. C. Konda, L. Loetgering, K. C. Zhou, S. Xu, A. R. Harvey, and R. Horstmeyer, "Fourier ptychography: current applications and future promises," Opt. Express **28**, 9603 (2020).
16. G. Zheng, C. Shen, S. Jiang, P. Song, and C. Yang, "Concept, implementations and applications of Fourier ptychography," Nat. Rev. Phys. **3**, 207–223 (2021).
17. C. Zheng, T. Wang, Z. Li, R. Sun, D. Yang, S. Wang, B. Ouyang, F. Liu, M. Xiang, Q. Hao, and S. Zhang, "Quantitative phase imaging based on Fourier ptychographic microscopy: advances, applications, and perspectives," Adv. Imaging **2**, 032001 (2025).
18. A. Greenbaum, W. Luo, T. W. Su, Z. Göröcs, L. Xue, S. O. Isikman, A. F. Coskun, O. Mudanyali, and A. Ozcan, "Imaging without lenses: Achievements and remaining challenges of wide-field on-chip microscopy," Nat. Methods **9**, 889–895 (2012).
19. A. Ozcan and E. McLeod, "Lensless Imaging and Sensing," Annu. Rev. Biomed. Eng. **18**, 77–102 (2016).
20. M. Rogalski, P. Arcab, E. Wdowiak, J. Á. Picazo-Bueno, V. Micó, M. Józwik, and M. Trusiak, "Hybrid Iterating-Averaging Low Photon Budget Gabor Holographic Microscopy," ACS Photonics **12**, 1771–1782 (2025).
21. Y. Wu and A. Ozcan, "Lensless digital holographic microscopy and its applications in biomedicine and environmental monitoring," Methods **136**, 4–16 (2018).
22. S. K. Vashist, P. B. Luppa, L. Y. Yeo, A. Ozcan, and J. H. T. Luong, "Emerging Technologies for Next-Generation Point-of-Care Testing," Trends Biotechnol. **33**, 692–705 (2015).
23. J. Zhang, J. Sun, Q. Chen, J. Li, and C. Zuo, "Adaptive pixel-super-resolved lensfree in-line digital holography for wide-field on-chip microscopy," Sci. Rep. **7**, 1–15 (2017).
24. Y. Gao and L. Cao, "Iterative projection meets sparsity regularization: towards practical single-shot quantitative phase imaging with in-line holography," Light Adv. Manuf. **4**, (2023).
25. Y. Rivenson, Y. Zhang, H. Günaydın, D. Teng, and A. Ozcan, "Phase recovery and holographic image reconstruction using deep learning in neural networks," Light Sci. Appl. **7**, 17141 (2018).
26. W. Bishara, U. Sikora, O. Mudanyali, T. W. Su, O. Yaglidere, S. Luckhart, and A. Ozcan, "Holographic pixel super-resolution in portable lensless on-chip microscopy using a fiber-optic array," Lab Chip **11**, 1276–1279 (2011).
27. W. Bishara, T.-W. Su, A. F. Coskun, and A. Ozcan, "Lensfree on-chip microscopy over a wide field-of-view using pixel super-resolution," Opt. Express **18**, 11181 (2010).
28. R. Cao, C. Shen, and C. Yang, "High-resolution, large field-of-view label-free imaging via aberration-corrected, closed-form complex field reconstruction," Nat. Commun. **15**, 1–8 (2024).
29. P. Gao and C. Yuan, "Resolution enhancement of digital holographic microscopy via synthetic aperture: a review," Light Adv. Manuf. **3**, 105–120 (2022).
30. K. Lee, H.-D. Kim, K. Kim, Y. Kim, T. R. Hillman, B. Min, and Y. Park, "Synthetic Fourier transform light scattering," Opt. Express **21**, 22453 (2013).
31. S. O. Isikman, W. Bishara, S. Mavandadi, F. W. Yu, S. Feng, R. Lau, and A. Ozcan, "Lens-free optical tomographic microscope with a large imaging volume on a chip," Proc. Natl. Acad. Sci. U. S. A. **108**, 7296–7301 (2011).
32. Y. Zhou, J. Yao, and L. V. Wang, "Optical clearing-aided photoacoustic microscopy with enhanced resolution and imaging depth," Opt. Lett. **38**, 2592 (2013).
33. Z. Shen, X. Guo, Y. Zhang, D. Li, and Y. He, "Enhancement of short coherence digital holographic microscopy by optical clearing," Biomed. Opt. Express **8**, 2036 (2017).
34. Y. Zhang, Y. Shin, K. Sung, S. Yang, H. Chen, H. Wang, D. Teng, Y. Rivenson, R. P. Kulkarni, and A. Ozcan, "3D imaging of optically cleared tissue using a simplified CLARITY method and on-chip microscopy," Sci. Adv. **3**, 1–11 (2017).
35. J. van Rooij and J. Kalkman, "Large-scale high-sensitivity optical diffraction tomography of zebrafish," Biomed. Opt. Express **10**, 1782 (2019).
36. D. Lee, M. Lee, H. Kwak, Y. S. Kim, J. Shim, J. H. Jung, W. Park, J.-H. Park, S. Lee, and Y. Park, "High-fidelity optical diffraction tomography of live organisms using iodixanol refractive index matching," Biomed. Opt. Express **13**, 6404 (2022).
37. Y. Liu, X. Yang, D. Zhu, R. Shi, and Q. Luo, "Optical clearing agents improve photoacoustic imaging in the optical diffusive regime," Opt. Lett. **38**, 4236 (2013).
38. W. Li, R. N. Germain, and M. Y. Gerner, "High-dimensional cell-level analysis of tissues with Ce3D multiplex volume imaging," Nat. Protoc. **14**, 1708–1733 (2019).
39. E. I. Chen, D. Cociorva, J. L. Norris, and J. R. Yates, "Optimization of Mass Spectrometry Compatible Surfactants for Shotgun Proteomics," J. Proteome Res. **6**, 2529–2538 (2007).
40. Z. Ou, Y.-S. Duh, N. J. Rommelfanger, C. H. C. Keck, S. Jiang, K. Brinson, S. Zhao, E. L. Schmidt, X. Wu, F. Yang, B. Cai, H. Cui, W. Qi, S. Wu, A. Tantry, R. Roth, J. Ding, X. Chen, J. A. Kaltschmidt, M. L.

Brongersma, and G. Hong, "Achieving optical transparency in live animals with absorbing molecules," Science (80-. ). **385**, eadm6869 (2024).
41. D. A. Miller, Y. Xu, R. Highland, V. T. Nguyen, W. J. Brown, G. Hong, J. Yao, and A. Wax, " Enhanced penetration depth in optical coherence tomography and photoacoustic microscopy in vivo enabled by absorbing dye molecules ," Optica **12**, 24 (2025).
42. A. Narawane, R. Trout, C. Viehland, A. N. Kuo, L. Vajzovic, A.-H. Dhalla, and C. A. Toth, "Optical clearing with tartrazine enables deep transscleral imaging with optical coherence tomography," J. Biomed. Opt. **29**, 1–7 (2024).
43. N. Yuan, S. Ragab, L. Chavez, V. Pandey, and X. Intes, "Evaluating Tartrazine as an Optical Clearing Agent for Fluorescence Lifetime Imaging," Opt. Lett. **50**, 7588–7591 (2025).
44. C. Jia, Z. Zhang, Y. Shen, W. Hou, J. Zhao, J. Luo, H. Chen, D. Qi, Y. Yao, L. Deng, H. Ma, Z. Sun, and S. Zhang, "Tartrazine-enabled optical clearing for in vivo optical resolution photoacoustic microscopy," Biomed. Opt. Express **16**, 2504 (2025).
45. M. Xu, B. Yang, S. Song, T. Xu, J. Yao, Y. Liu, Y. Cui, and Y. Zhang, "Multi-wavelength photoacoustic microscopy enhanced by the high-sensitivity probe and reversible tissue transparent molecules," Photonics Res. **13**, 2757 (2025).
46. T. Zuo, C. Tao, and X. Liu, "Absorbing molecules as optical clearing agents improve the resolution and sensitivity of photoacoustic microscopy," Opt. Lett. **50**, 2282 (2025).
47. Q. Xu and J. Zhu, "Glycerol concentration sensor based on the MIM waveguide structure," Front. Phys. **10**, 1–9 (2022).
48. U. Leischner, M. Reifarth, M. S. Brill, F. Schmitt, S. Hoeppener, D. Unnersjö Jess, H. Brismar, U. S. Schubert, and R. Heintzmann, "PRIAMOS: A technique for mixing embedding media for freely adjusting pH value and refractive index (RI) for optical clearing (OC) of whole tissue samples," J. Microsc. 1–13 (2025).
49. Y. Kaneoke, T. Donishi, J. Iwatani, S. Ukai, K. Shinosaki, and M. Terada, "Variance and Autocorrelation of the Spontaneous Slow Brain Activity," PLoS One **7**, e38131 (2012).
50. G. E. Lang, P. S. Stewart, D. Vella, S. L. Waters, and A. Goriely, "Is the Donnan effect sufficient to explain swelling in brain tissue slices?," J. R. Soc. Interface **11**, (2014).
51. A. R. Guerra, L. R. Oliveira, G. O. Rodrigues, M. R. Pinheiro, M. I. Carvalho, V. V. Tuchin, and L. M. Oliveira, "Assessment of tartrazine diffusion properties in skeletal muscle," IEEE J. Sel. Top. Quantum Electron. **32**, 1–8 (2025).
52. A. Jaafar, M. E. Darvin, V. V. Tuchin, and M. Veres, "Confocal Raman Micro-Spectroscopy for Discrimination of Glycerol Diffusivity in Ex Vivo Porcine Dura Mater," Life **12**, (2022).
53. R. D. Bülow and P. Boor, "Extracellular Matrix in Kidney Fibrosis: More Than Just a Scaffold," J. Histochem. Cytochem. **67**, 643–661 (2019).
54. C. E. McQuitty, R. Williams, S. Chokshi, and L. Urbani, "Immunomodulatory Role of the Extracellular Matrix Within the Liver Disease Microenvironment," Front. Immunol. **11**, 1–33 (2020).
55. M. Magzoub, S. Jin, and A. S. Verkman, "Enhanced macromolecule diffusion deep in tumors after enzymatic digestion of extracellular matrix collagen and its associated proteoglycan decorin," FASEB J. **22**, 276–284 (2008).
56. K. Moulton, F. Lovell, E. Williams, P. Ryan, D. C. Lay, D. Jansen, and S. Willard, "Use of glycerol as an optical clearing agent for enhancing photonic transference and detection of Salmonella typhimurium through porcine skin," J. Biomed. Opt. **11**, 054027 (2006).
57. W. Li, R. N. Germain, and M. Y. Gerner, "Multiplex, quantitative cellular analysis in large tissue volumes with clearing-enhanced 3D microscopy (Ce3D)," Proc. Natl. Acad. Sci. U. S. A. **114**, E7321–E7330 (2017).
58. L. Tian, X. Li, K. Ramchandran, and L. Waller, "Multiplexed coded illumination for Fourier Ptychography with an LED array microscope," Biomed. Opt. Express **5**, 2376 (2014).
59. K. Niedziela, M. Rogalski, P. Arcab, J. Winnik, P. Zdańkowski, and M. Trusiak, "Pixel Super Resolution with Axial Scanning in Lensless Digital In-line Holographic Microscopy," Photonics Lett. Pol. **16**, 76–78 (2024).

# Supplementary Document 1: The mixture of glycerin with tartrazine: a solution to reversibly increase tissue transparency for *in vitro* quantitative phase imaging

MIKOŁAJ KRYSA[1,*], ANNA CHWASTOWICZ[2,3], MAŁGORZATA LENARCIK[4,5], PAWEŁ MATRYBA[2] PIOTR ZDAŃKOWSKI[1] AND MACIEJ TRUSIAK[1,**]

[1]*Institute of Micromechanics and Photonics, Warsaw University of Technology, Warsaw, Poland*
[2] *Department of Immunology, Medical University of Warsaw, Warsaw, Poland*
[3] *Laboratory of Neurobiology, Nencki Institute of Experimental Biology of the Polish Academy of Sciences, Warsaw, Poland*
[4]*Department of Pathology, Maria Sklodowska-Curie National Research Institute of Oncology, Warsaw, Poland*
[5]*Department of Gastroenterology, Hepatology and Clinical Oncology, Centre of Postgraduate Medical Education, Warsaw, Poland*
*\*mikolaj.krysa@pw.edu.pl*
*\*\*maciej.trusiak@pw.edu.pl*

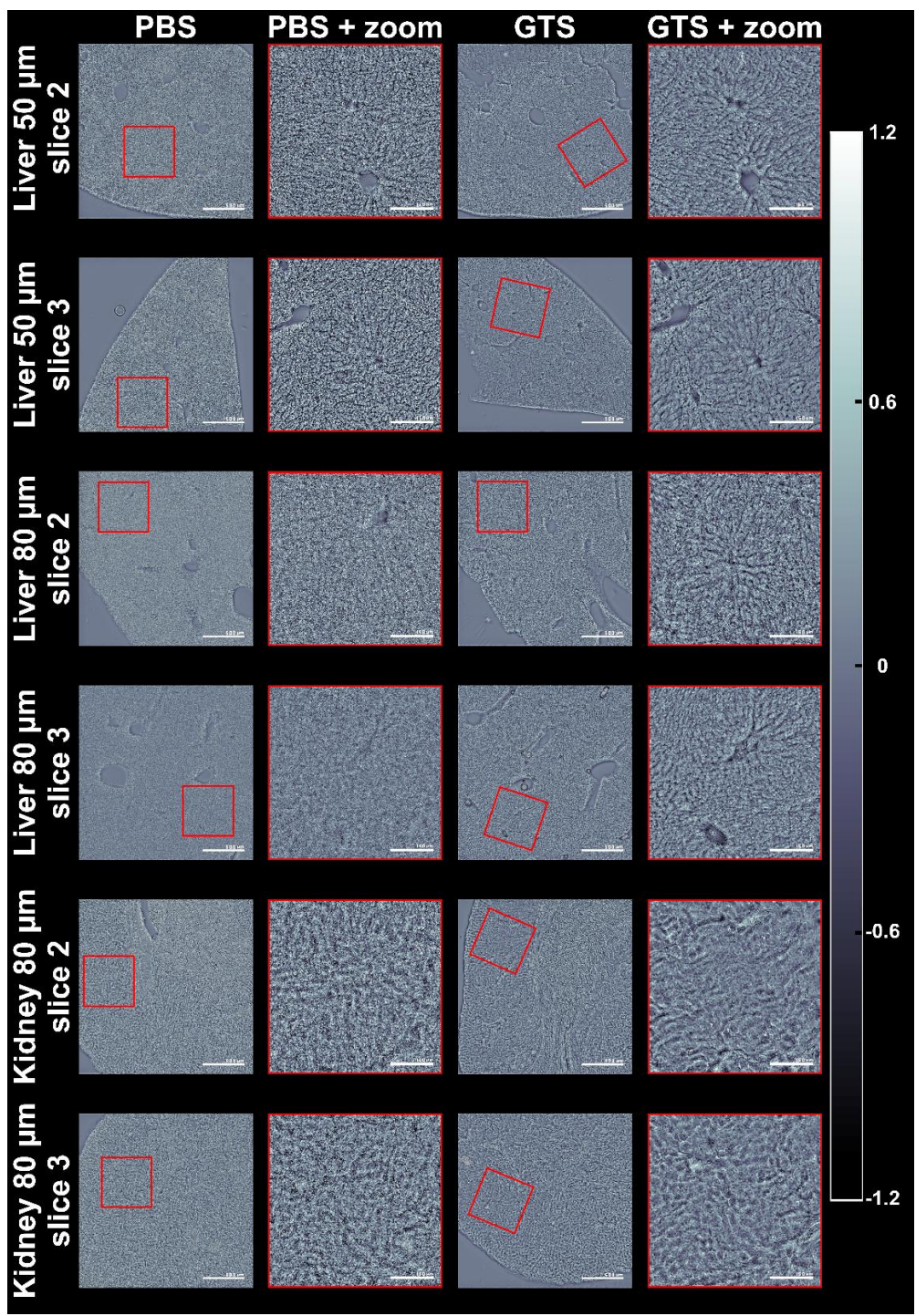

Fig. S1. The reconstructed phase images of additional liver (50 μm and 80 μm thick) and kidney (80 μm thick) slices obtained with Fourier Ptychographic Microscopy. The images show the reconstructed phase from the samples in PBS and in Glycerol and Tartrazine Solution (GTS). The red rectangles mark the zoomed region. The same areas are shown in the zoomed

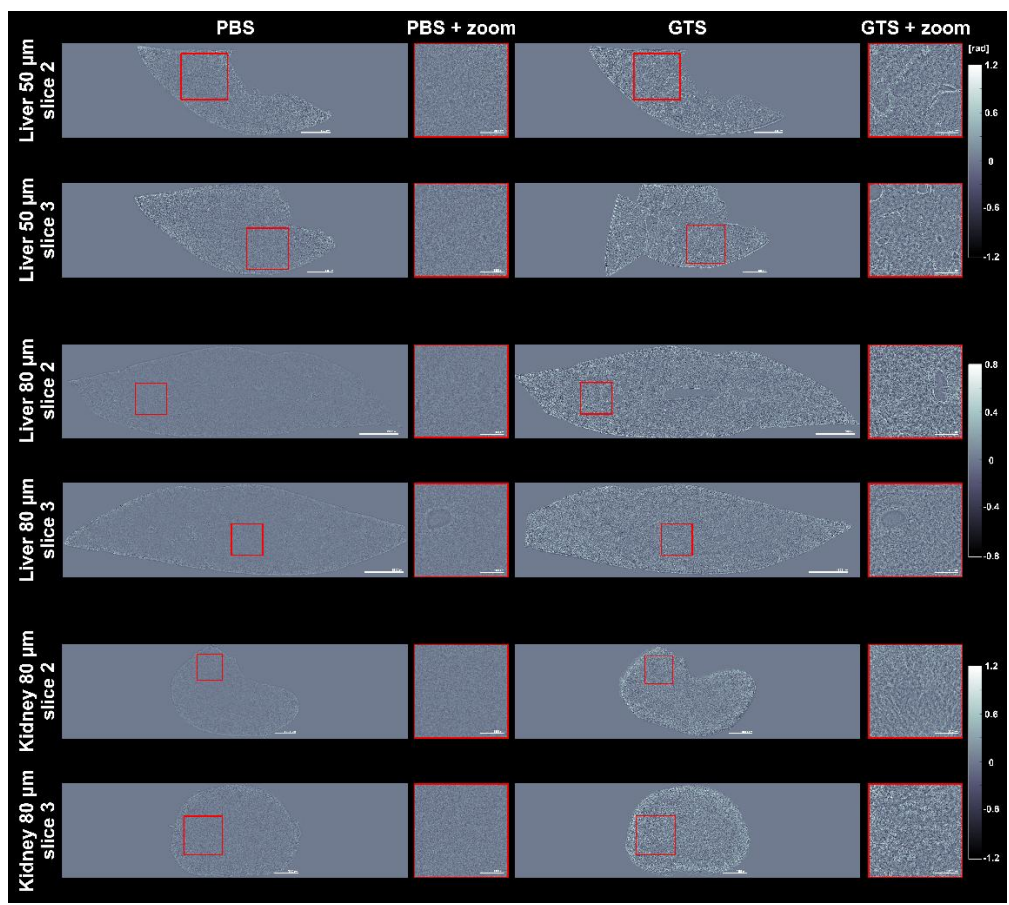

Fig. S2. The reconstructed phase images of additional liver (50 µm and 80 µm thick) and kidney (80 µm thick) slices obtained with Lensless Digital Holographic Microscopy with Pixel Super Resolution. The images show the reconstructed phase from the samples in PBS and in Glycerol and Tartrazine Solution (GTS). The red rectangles mark the zoomed region. The same areas are shown in the zoomed images. The color bar shows the phase values. The analysis in PBS and GTS was carried out on the same slice.

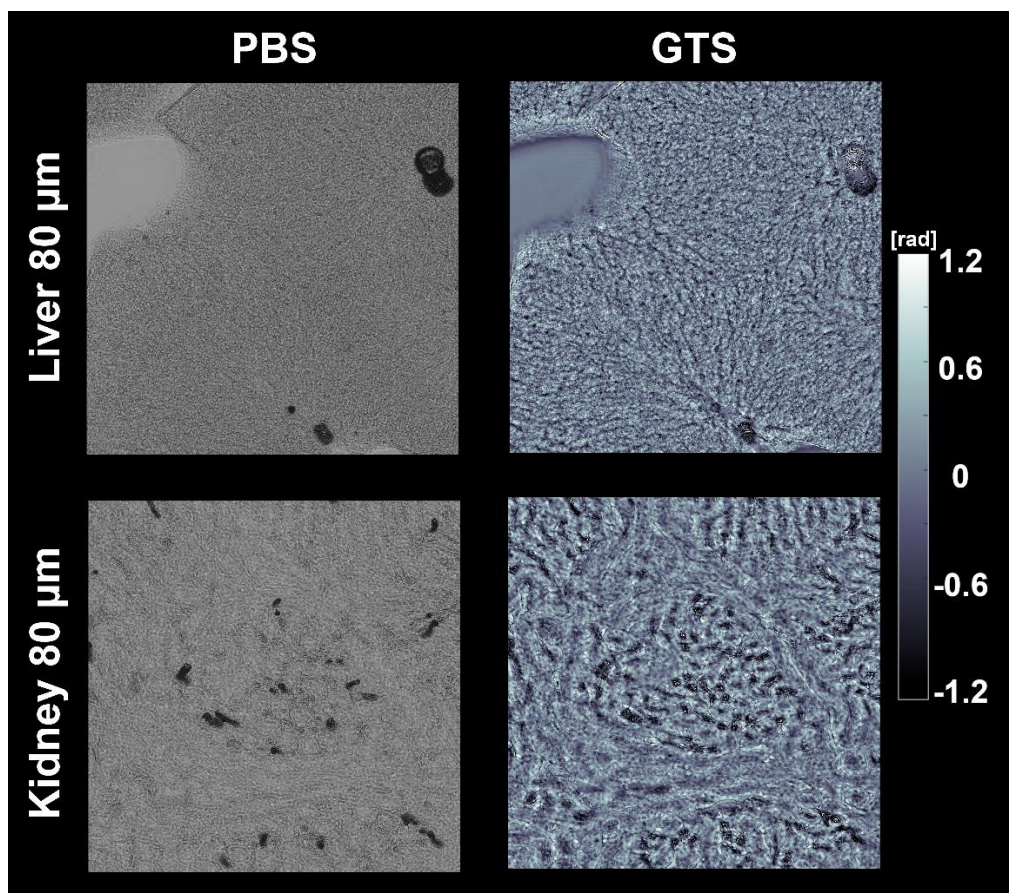

Fig. S3. The zoomed, reconstructed amplitude and phase images of 80 μm thick liver and kidney slices obtained with Fourier Ptychographic Microscopy. The images show the samples in Glycerol and Tartrazine Solution (GTS). The red rectangles mark the zoomed region. The color bar shows the phase values. The lowest phase values overlay with the regions with the most absorption shown in the amplitude image.

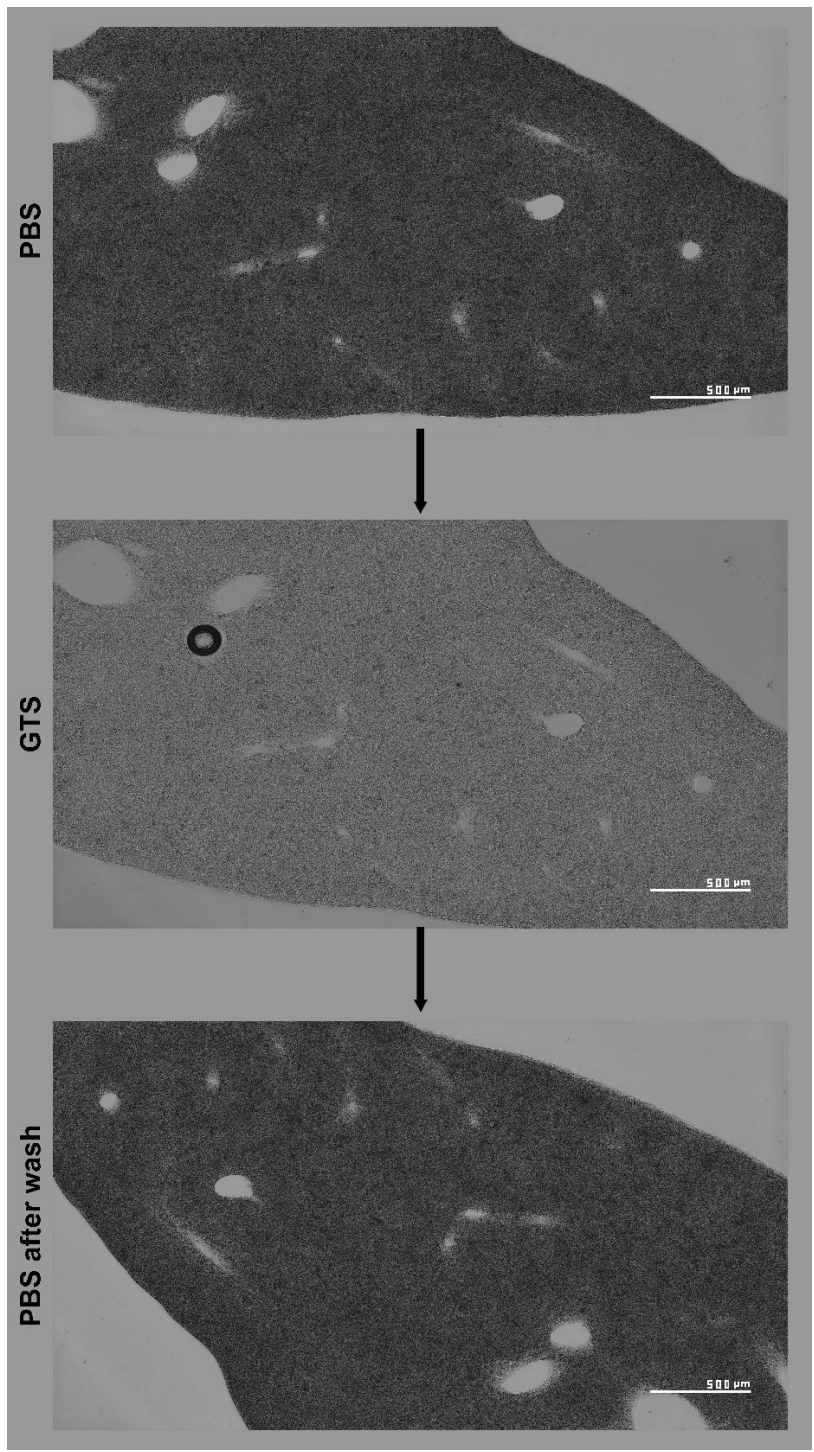

Fig. S4. The presentation of the reversible GTS clearing effect on the amplitude images. The images are reconstructed from the amplitude of 80 μm thick liver slices obtained with Fourier Ptychographic Microscopy. The sample was first submerged in PBS, then in GTS, and subsequently washed for 15 minutes with PBS. It was imaged after each of these steps.